*arXiv:0706.3985v2 [stat.ME] 13 Dec 2007*

# DISTRIBUTIONS ASSOCIATED WITH GENERAL RUNS AND PATTERNS IN HIDDEN MARKOV MODELS

By John A. D. Aston and Donald E. K. Martin

*Academia Sinica, and United States Census Bureau, Howard University and North Carolina State University*

This paper gives a method for computing distributions associated with patterns in the state sequence of a hidden Markov model, conditional on observing all or part of the observation sequence. Probabilities are computed for very general classes of patterns (competing patterns and generalized later patterns), and thus, the theory includes as special cases results for a large class of problems that have wide application. The unobserved state sequence is assumed to be Markovian with a general order of dependence. An auxiliary Markov chain is associated with the state sequence and is used to simplify the computations. Two examples are given to illustrate the use of the methodology. Whereas the first application is more to illustrate the basic steps in applying the theory, the second is a more detailed application to DNA sequences, and shows that the methods can be adapted to include restrictions related to biological knowledge.

**1. Introduction.** Hidden Markov models (HMMs) provide a rich structure for use in a wide range of statistical applications. As examples, they serve as models in speech recognition [Rabiner (1989)], image processing [Li and Gray (2000)], DNA sequence analysis [Durbin, Eddy, Krogh and Mitchison (1998) and Koski (2001)], DNA microarray time course analysis [Yuan and Kendziorski (2006)] and econometrics [Hamilton (1989) and Sims and Zha (2006)], to name just a few. HMMs essentially specify two structures, an underlying model for the unobserved state of the system, and one for the observations, conditional on the unobserved states. Thus, HMMs are a sub-class of state space models [Harvey (1989)], but have the restriction that the models for the hidden states are defined on finite dimensional spaces.

HMMs have been studied extensively, especially for the case where the hidden sequence is first-order Markovian (a Markov chain); see, for example, Rabiner (1989). Higher-order HMMs are less frequently used, but are gaining in popularity, especially in areas such as bioinformatics [Krogh (1997) and Ching, Ng and Fung (2003)]. For practical purposes, three fundamental problems associated with first-order HMMs have been examined thoroughly and solved [Rabiner (1989)]: (1) the efficient computation of the likelihood of the sequence of observations given the HMM [Baum and Eagon (1967)]; (2) the determination of a best sequence of underlying states to maximize the likelihood of the observation sequence [Viterbi (1967)]; and (3) the adjustment of model parameters to maximize the likelihood of the observations [the Baum–Welch algorithm; see Baum and Eagon (1967)].

However, little is known about probabilities for patterns or collections of patterns (also known as words or motifs, resp.) in heterogeneous sequences such as those of HMMs [Reinert, Schbath and Waterman (2005)]. In this paper a fourth problem that is becoming increasingly more important for applications such as bioinformatics and data mining is considered: the probability that a pattern has occurred or will occur in the hidden state sequence of an HMM.

Currently, inference on patterns in the hidden state sequence of an HMM usually proceeds as follows. The HMM is determined and the Viterbi algorithm is used to find the most probable state sequence among all possible ones, conditional on the observations. This state sequence is then treated as if it is "deterministically correct" and patterns are found by examining it. However, the conditional distribution (given the observations) of patterns over all state sequences is more relevant. If, for example, the number of genes present in a DNA sequence is of interest and the Viterbi sequence of an HMM is used [as in methods based on Krogh, Mian and Haussler (1994)], then counting genes from the Viterbi sequence cannot be guaranteed to even give a good estimate of the number of genes. This is because there could be gene counts that correspond to many state sequences, and when accumulating probabilities over those sequences, one could find that those counts are much more likely than the count corresponding to the Viterbi sequence. This could especially be true if there are many different sequences all with likelihood close to that of the Viterbi sequence. If a single choice of gene count is needed, then the mean of the conditional distribution over state sequences, given the observations, would seem to be a more reasonable choice. Thus, a method to compute pattern distributions in state sequences modeled as HMMs would be helpful.

In this paper a computational method for finding such pattern distributions is developed, and waiting time probabilities for patterns under the general framework given in Aston and Martin (2005) are extended to the state sequence of HMMs. The probabilities are computed under the paradigm that $T$ observations of the output sequence of the HMM have been realized, and the generation process is either complete or set to continue. Waiting time probabilities are then computed for patterns in the unobserved state sequence up to a time $T^* \geq T$. Note that waiting time probabilities for patterns in the observations of an HMM when no data has yet been observed (i.e., $T = 0$), a case that relates directly to computations in the standard case of Markovian sequences with no "hidden" states, was dealt with in Cheung (2004).

The methodology of this paper will be applied to two examples, but one application will be studied in detail, that of $CpG$ Island analysis in DNA sequences. A $CpG$ island is a short segment of DNA in which the frequency of $CG$ pairs is higher than in other regions. The "$p$" indicates that $C$ and $G$ are connected by a phosphodiester bond. The $C$ in a $CG$ pair is often modified by methylation, and if that happens, there is a relatively high chance that it will mutate to a $T$, and thus, $CG$ pairs are under-represented in DNA sequences. Upstream from a gene, the methylation process is suppressed in a short region of length 100–5,000 nucleotides, known as $CpG$ islands [Bird (1987)]. The underlying nucleotide generating sequence can be modeled as two different systems, one for when the sequence is in a $CpG$ island, and one for when it is not.

As $CpG$ islands can be especially useful for identifying genes in human DNA [Takai and Jones (2002)], different methods have been developed for their detection. Two distinct methods for $CpG$ island analysis are common. First, a windowed segment of the DNA sequence is taken and the number of $CG$ pairs within it is counted. If the frequency of $CG$ pairs is higher than some predetermined threshold, then the segment is defined to be within a $CpG$ island [Takai and Jones (2002)]. Popular software such as $CpG$ Island Searcher [Takai and Jones (2003)] proceeds along these lines. However, these types of methods have been criticized due to the need for the predetermination of thresholds and the variability in the results that arise depending on the choice [Durbin, Eddy, Krogh and Mitchison (1998) and Saxonov, Berg and Brutlag (2006)].

A second method of determining islands, which overcomes the need for thresholds, is to use HMMs for the analysis [Durbin, Eddy, Krogh and Mitchison (1998)]. Software is readily available to implement these HMM-based methods for $CpG$ island analysis, for example, Guéguen (2005). The Viterbi algorithm is used to segment the sequence and analysis then proceeds as indicated above using the "deterministic" Viterbi sequence. However, it will be shown that by computing other distributions such as the distribution of the number of $CpG$ islands and the distribution of $CpG$ island lengths, additional biologically useful information may be obtained. Specific examples will also be given to illustrate that the number of $CpG$ islands found by the Viterbi algorithm does not necessarily represent the most probable number in sets of human DNA data, and that the length of the islands found by the Viterbi algorithm can be significantly longer than expected.

The paper is organized as follows. In the next section some background material on hidden Markov models is given, including theorems establishing the forward and backward algorithms in the general $m$th order case analogous to those when $m = 1$. Also included in the section is information on the types of patterns that the theory covers, the counting techniques that may be used, and the methodology that is used to compute probabilities. In Section 3 the results of the paper are given, namely, the lemmas and theorems required to calculate distributions associated with patterns in the state sequence of an HMM. Section 4 contains two examples. A simple application to geological data is used to demonstrate the ideas, and then the analysis of the system just discussed, that of $CpG$ islands in DNA sequences, is undertaken. It will be shown that a variety of distributions of interest may be calculated, including distributions under biological constraints such as minimum island lengths and minimal separation between islands. Section 5 is a discussion.

## 2. Notation and preliminaries.

2.1. *Hidden Markov models.* Let $\{Y_t\}_{t=1}^T$ be the observations of a higher-order HMM, generated from an underlying unobserved state sequence $\{X_t\}_{t=-m+1}^{T^*}$ that is assumed to be a Markovian sequence of a general order $m$. (Frequently sequences will be denoted using $\{\cdot\}$ notation without giving the specific index set, e.g., $\{Y_t\}$ and $\{X_t\}$ for the observation and state sequences above, not to be confused with generic sequence values $Y_t$ and $X_t$.) While there are observations $Y_1, \ldots, Y_T$ to index $T$ (that is usually referred to as "time" in this paper, but could simply be the sequence index), it is assumed that the system is set to continue until a time $T^* \geq T$, allowing for the consideration of systems where only partial observation sequences have been realized. The states $X_{-m+1}, \ldots, X_0$ are defined to allow the initialization of the system.

Denote the state space of $\{X_t\}$ as $S_X$, and similarly, let $S_Y$ be the state space of the observation sequence. $S_Y$ is assumed to be finite for simplicity, although this assumption may be relaxed without causing much difficulty. The initial distribution of the state sequence $\{X_t\}$ is denoted by

$$(1) \qquad \pi(X_{-m+1}, \ldots, X_0),$$

and the time-homogeneous transition probabilities are denoted by

$$(2) \qquad p(X_t | X_{t-m}, \ldots, X_{t-1}).$$

An alternative setup is to place the initial distribution on $X_1, \ldots, X_m$, which will lead to analogous results if not identical ones.

Also denote the observation probabilities, conditioned on the state $X_t$, by $\gamma_t(Y_t|X_t)$, $t = 1, \ldots, T$. The assumption is made that, conditional on $X_{-m+1}, \ldots, X_{T^*}$, the $Y_t$'s are independent, and that the conditional distribution of each $Y_t$ given $X_{-m+1}, \ldots, X_{T^*}$ depends only on $X_t$. These assumptions were taken as fundamental to HMMs in MacDonald and Zucchini (1997) with $T^* = T$, that is, the latter authors were only concerned with times up to time $T$ and not $T^*$. The assumptions imply that the $Y_t$'s are independent given $X_{-m+1}, \ldots, X_T$ as well.

In an HMM of order $m$, conditional probabilities for the hidden state $X_t$ given $X_{t-m}, \ldots, X_{t-1}$ are independent of the more distant past. Usually the assumption that $m = 1$ is made, however, in this paper, $m$ may take on general positive integer values. Rabiner (1989) considered HMMs as both a generator of the observations, and as a model for how a given observation sequence $Y^{(T)}$ was generated by an appropriate HMM. It should be noted that taking either the MacDonald and Zucchini (1997) or Rabiner (1989) viewpoints toward HMMs leads to computationally identical answers.

The forward and backward algorithms [Baum and Eagon (1967)] are helpful for computing probabilities associated with the HMM. These algorithms are well known for the first-order HMM case; here versions are stated for general orders of dependence. While it is true that a higher-order HMM may be posed as a first-order HMM, the forward and backward algorithms that follow can be more efficient if the higher-order model is used. The proofs of these algorithms follow in a straightforward manner from the proofs of the original forward and backward algorithms and are therefore omitted.

For notational purposes, let $Y^{(j)} \equiv Y_1, \ldots, Y_j$, $j \geq 1$, $X^{(l)} \equiv X_{-m+1}, \ldots, X_l$, $l \geq (-m+1)$ and $\tilde{X}_t \equiv (X_{t-m+1}, \ldots, X_t)$, $t = 0, 1, \ldots, T^*$. Now define the "forward" variables by $\alpha_0(\tilde{X}_0) \equiv \pi(\tilde{X}_0)$, and

$$(3) \qquad\qquad \alpha_t(\tilde{X}_t) \equiv P(Y^{(t)}, \tilde{X}_t), \qquad t = 1, \ldots, T.$$

THEOREM 2.1. *For an $m$th order HMM, the forward variables may be computed recursively, and the probability of the observation sequence $Y^{(T)}$ may be computed from them, using the following:*

(i) *Initialization:*

$$(4) \qquad\qquad \alpha_0(\tilde{X}_0) = \pi(\tilde{X}_0).$$

(ii) *Recursion: For $t = 1, \ldots, T$,*

$$(5) \qquad \alpha_t(\tilde{X}_t) = \sum_{X_{t-m}} \alpha_{t-1}(\tilde{X}_{t-1}) p(X_t|\tilde{X}_{t-1}) \gamma_t(Y_t|X_t).$$

(iii) *Termination:*

$$(6) \qquad\qquad P(Y^{(T)}) = \sum_{\tilde{X}_T} \alpha_T(\tilde{X}_T).$$

Also define the "backward" variables by $\beta_T(\tilde{X}_T) = 1$ for all $\tilde{X}_T$, and

$$(7) \qquad \beta_t(\tilde{X}_t) \equiv P(Y_{t+1}, \ldots, Y_T | \tilde{X}_t), \qquad 0 \le t \le T - 1.$$

THEOREM 2.2. *For an $m$th order HMM, the backward variables may be computed recursively using the following:*

(i) *Initialization:*

$$(8) \qquad \beta_T(\tilde{X}_T) = 1 \qquad \text{for all } \tilde{X}_T.$$

(ii) *Recursion: For $t = T - 1, T - 2, \ldots, 1, 0$,*

$$(9) \qquad \beta_t(\tilde{X}_t) = \sum_{X_{t+1}} p(X_{t+1} | \tilde{X}_t) \gamma_{t+1}(Y_{t+1} | X_{t+1}) \beta_{t+1}(\tilde{X}_{t+1}).$$

2.2. *Notation for patterns.* A *simple pattern* $\Lambda_i$ refers to a specified sequence of symbols of $S_X$, where the symbols are allowed to be repeated. A *compound pattern* $\Lambda$ is the union of simple patterns, that is, $\Lambda = \bigcup_{i=1}^{\eta} \Lambda_i$, where the lengths of the simple patterns, $\Lambda_i$ may vary, $\Lambda_a \cup \Lambda_b$ denotes the occurrence of pattern $\Lambda_a$ or pattern $\Lambda_b$, and the integer $\eta \ge 1$.

Consider now a system $\Lambda^{(1)}, \ldots, \Lambda^{(c)}$ of $c$ compound patterns, $c \ge 1$, with corresponding numbers $r_1, \ldots, r_c$, where $r_j$ denotes the required number of occurrences of compound pattern $\Lambda^{(j)}$. If the waiting time of interest is the time until the first occurrence of one of the compound patterns its specified number of times, $\Lambda^{(1)}, \ldots, \Lambda^{(c)}$ are called *competing patterns* [Aston and Martin (2005)]. If all of the patterns must occur their specified number of times, the system is called *generalized later patterns* [Martin and Aston (2007)].

When $c = 1$, results for competing and generalized later patterns reduce to those for compound patterns. On the other hand, if $r_j = 1$ for all $j$ and, in addition, each of the competing or generalized later patterns consists of just one simple pattern, then the respective waiting time distributions reduce to *sooner* and *later waiting time distributions* as defined in the literature; for example, see Balakrishnan and Koutras (2002).

A special simple pattern is a *run* of length $k$, where run is used in a loose sense that includes consecutive occurrences of any pattern such as a single symbol or alternating symbols, as in examples of Section 4, when the length distribution of the longest run is considered. A run of length $k$ will be denoted by $\Lambda(k)$ and its waiting time by $W(\Lambda(k))$, where the type of run will be understood by the context.

2.3. *Methods of counting.* Two distinct methods of counting patterns are possible, though much of the implementation will not change dramatically between the two cases. The first method is that of *nonoverlapping* counting. With that method, when a pattern occurs, the counting process re-starts from that point, and any partially completed pattern cannot be finished. Nonoverlapping counting can be system wide (SWNO), or it can be restricted to within compound patterns (WPNO), where counting only starts over for simple patterns within the same compound pattern as the one that has just occurred [Aston and Martin (2005)]. The second counting method is that of *overlapping* counting, where partially completed patterns may be finished at any time, regardless of whether another pattern has been completed after the partially completed pattern starts but before it is completed.

To illustrate, let $\Lambda^{(1)} = \Lambda(5) = 11111$ and $\Lambda^{(2)} = 1011 \cup 00$. For the realization

$$(10) \qquad (X_1, \ldots, X_{15}) = 011011111101100,$$

using SWNO counting, there are two occurrences of 1011 and one of 00 (and thus, three occurrences of $\Lambda^{(2)}$), but no occurrences of $\Lambda^{(1)}$ are counted. Although there are six consecutive ones in the realization, only four of them occur after the first completion of 1011 (which re-starts counting), and thus, 11111 is not counted. With WPNO counting, in addition to the occurrences mentioned above, $\Lambda^{(1)}$ is counted once, since now the first occurrence of 1011 doesn't cause a re-starting of counting for 11111, which is in a different compound pattern. Finally, with overlapping counting, $\Lambda^{(1)}$ occurs twice (the two overlapping occurrences of 11111), in addition to the three occurrences of $\Lambda^{(2)}$.

2.4. *Method of computation.* The use of an auxiliary Markov process to aid with the computation of probabilities has been used for more than fifty years. As an example, Cox (1955) used supplementary variables to obtain a Markov process that simplified the computation of waiting time probabilities, and noted that the use of supplementary variables in that manner was a well-known technique at the time. In a recent paper concerning Markov Modulated Poisson Processes [Hwang, Choi and Kim (2002)], waiting time distributions were computed in a similar manner. Fu and Koutras (1994) used auxiliary Markov chains, and developed an elegant framework [calling it *Finite Markov Chain Imbedding* (FMCI)] for handling distributions associated with runs and patterns in discrete time Markov processes.

In the FMCI method, a Markov chain $\{Z_t\}$ is developed such that there is a one-to-one correspondence between classes of its states and those of $\{X_t\}$. The Markov chain $\{Z_t\}$ facilitates the computation of waiting time distributions for patterns, since an absorbing state for $\{Z_t\}$ can be defined to correspond to the occurrence of the pattern of interest. The desired probability may then be computed by determining the probability that $\{Z_t\}$ lies in the corresponding absorbing state, which is computed using basic properties of Markov chains. Aston and Martin (2005) and Martin and Aston (2007) used the FMCI approach to compute waiting time distributions for competing and generalized later patterns, respectively, in higher-order Markovian sequences, and that method is also used in this paper.

As is the case when converting an $m$th order Markovian sequence into a first-order one through the use of $m$-tuples, the $m$-tuple $X_{t-m+1}, \ldots, X_t$ must be a part of the state of $Z_t$. To deal with competing and generalized later patterns in general, information on the number of compound pattern occurrences to date and current progress into the simple patterns of the system must also be part of the state of $Z_t$, so that probabilities for $Z_{t+1}$ may be determined based only on $Z_t$, and not on earlier time points.

Once the state space $S_Z$ of $\{Z_t\}$ is determined, entries of the associated transition probability matrix are determined by first noting the possible transitions for $\{X_t\}$ (and the implied transitions for $\{Z_t\}$), and then assigning the appropriate transition probabilities. The initial distribution associated with $Z_0$ is determined by mapping $m$-tuples $(x_{-m+1}, \ldots, x_0)$ into corresponding $Z_0$ values, and then assigning the corresponding probabilities $\pi(X_{-m+1}, \ldots, X_0)$. All other initial states will have probability zero. The formation of the state space $S_Z$ and the determination of its transition probability matrix and initial distribution will be illustrated in the examples.

In the next section it is shown that the results of the latter two references may be extended to derive distributions associated with patterns in the state sequence of an HMM. The fundamental difference between distributional problems dealt with in this paper and those for regular Markovian sequences is that in the present setup the observed data up to time $T$ must be accounted for, whereas regular Markovian sequences are equivalent to the HMM case where no observed data is available ($T = 0$).

## 3. Distributions associated with patterns in HMM states.
In this section results on distributions associated with pattern occurrences in the state sequence of an HMM of a general order are presented.

To apply the methods described above to the underlying state sequence $\{X_t\}$ of an HMM, it is necessary to know the transition probabilities of $\{X_t\}$ conditional on the observed values $Y^{(T)}$. These probabilities are given by the following lemma. A theorem giving a formula for computing the waiting time probabilities then follows, along with results that may be obtained from that theorem.

LEMMA 3.1.  *For an $m$th order HMM, if $1 \le t \le T$, then*

$$(11) \qquad P(X_t | X^{(t-1)}, Y^{(T)}) = \frac{\beta_t(\tilde{X}_t)}{\beta_{t-1}(\tilde{X}_{t-1})} p(X_t | \tilde{X}_{t-1}) \gamma_t(Y_t | X_t).$$

*If $T < t \le T^*$, then*

$$(12) \qquad P(X_t | X^{(t-1)}, Y^{(T)}) = p(X_t | \tilde{X}_{t-1}).$$

The special case of (11) with $m = 1$ was given in Lindgren (1978), and the proof follows in a similar manner and is therefore omitted.

THEOREM 3.2 (Waiting time distribution for patterns in HMMs). *Consider an $m$th order HMM with state sequence $\{X_t\}_{t=-m+1}^{T^*}$ and observation sequence $\{Y_t\}_{t=1}^{T}$. If $\{Z_t\}_{t=0}^{T^*}$ is a Markov chain such that, conditioned on $Y^{(T)}$, the desired pattern has occurred by time $t$ if and only if $Z_t$ lies in a corresponding absorbing state $\Gamma \in S_Z$, then, conditional on $Y^{(T)}$, the waiting time distributions for competing and generalized later patterns may be computed through $\{Z_t\}$ by the equation*

$$(13) \qquad P(Z_t = \Gamma) = \psi_0 \left( \prod_{j=1}^{t} M_j \right) U(\Gamma),$$

*where $M_j$ is the transition probability matrix for transitions from $Z_{j-1}$ and $Z_j$, $\psi_0$ is a row vector holding the initial distribution of $Z_0$, and $U(\Gamma)$ is a column vector with a one in the location corresponding to absorbing state $\Gamma$, and zeros elsewhere.*

PROOF. By assumption, the pattern of interest has occurred by time $t$ if and only if $Z_t = \Gamma$. Equation (13) then follows using the well-known Chapman–Kolmogorov equations. $\square$

An equation similar to (13) is also used in the aforementioned research that uses supplementary variables to calculate waiting time distributions. The entries of the matrices $M_t$ are based on transitions for the underlying state sequence $\{X_t\}$ conditioned on $Y^{(T)}$, and thus are determined using Lemma 3.1.

The distribution of the length of the longest run of symbols may be computed through waiting time probabilities. The relationship between the probability mass function of $R(t)$, the length of the longest run that has occurred by time $t$, and waiting time probabilities for runs is given by

$$(14) \qquad \begin{aligned} P(R(t) = k) &= P(R(t) \ge k) - P(R(t) \ge k+1) \\ &= P(W\{\Lambda(k)\} \le t) - P(W\{\Lambda(k+1)\} \le t). \end{aligned}$$

The distribution of the number of runs of length at least $k$ may also be computed remarkably easily through associating Markov chains $\{Z_t^{(i)}\}$, $i = 1, \ldots, r$ with each occurrence ($r$ being the maximum number of occurrences of interest), and then using a renewal argument. An absorbing state $\Gamma$ is included in $S_Z$ to denote the occurrence of a run of length at least $k$. Also placed in $S_Z$ are "continuation" states $(s, \rightarrow)$ that indicate that the $(i-1)$st occurrence of the run is still in progress in $\{Z_t^{(i)}\}$, where $s$ denotes an $m$-tuple of the last $m$ state values that either led to the occurrence or to the continuation of the run (as with the absorbing state $\Gamma$, a common set of continuation states are used by the various chains $\{Z_t^{(i)}\}$, although the exact interpretation of all the states in $S_Z$ depends on which chain is being considered). Entries are added into the transition matrix $M_t$ corresponding to transition probabilities from the continuation states into the other states. The continuation states either transition to one of the continuation states, which would indicate the run is still in progress, or to one of the other states in $S_Z$ when the run has ended. The probabilities for the continuation state transitions are determined in an identical way to the other transition probabilities in $M_t$. No states in $S_Z$ transition into the continuation states other than the continuation states themselves. Otherwise, in this algorithm, there are no changes to $M_t$.

The first Markov chain $\{Z_t^{(1)}\}$ follows the movements of $\{Z_t\}$ up to the first occurrence of a run of length $k$, and with that occurrence the chain enters state $\Gamma$. The chain is initialized through $\psi_0$, with zeroes added for the continuation states. Subsequent chains $\{Z_t^{(i)}\}$, $i = 2, \ldots, r$, are initialized in the "continuation" states since a pattern of length $k$ has just occurred, and track the movements of $\{Z_t\}$ between the $(i-1)$st and $i$th run occurrences. Each chain enters the absorbing state once the run of the desired length occurs.

In this algorithm, the $1 \times |S_Z|$ row vector $\psi_t = \psi_0(\prod_{j=1}^t M_j)$ of (13), which gives the probability of the chain lying in its various states at time $t \geq 1$, is now replaced by an $r \times |S_Z|$ matrix $\Psi_t$ ($|B|$ denotes the number of elements of the set $B$). The matrix elements $\Psi_t(i, j)$ give the joint probabilities of there having been $(i-1)$ run occurrences of length at least $k$, and the chain $\{Z_t^{(i)}\}$ lying in state $j$. These probabilities must be incremented at time $t$ by adding the probability that the $(i-1)$st run occurs at that time. At time $t = 0$, the first row of $\Psi_0$ consists of $\psi_0$, with zeroes added to account for the continuation states, as mentioned above. Rows $2, \ldots, r$, respectively, corresponding to the chains $\{Z_t^{(i)}\}$, $i = 2, \ldots, r$, consist of all zeroes, since there is zero probability that there has been an occurrence of a run of length $k$ or longer at that time. The algorithm is given in the following theorem.

THEOREM 3.3 (Number of run occurrences of at least a certain length in HMMs).   *For an $m$th order HMM and auxiliary Markov chain $\{Z_t\}_{t=0}^{T^*}$ as in Theorem 3.2, let $\{Z_t^{(i)}\}_{t=0}^{T^*}$ be the corresponding chains that track movements of $\{Z_t\}_{t=0}^{T^*}$ between the $(i-1)$st and $i$th run of length at least $k$. Then* *conditional on $Y^{(T)}$, the distribution of the number of runs of length at least $k$ may be computed through*

$$(15) \qquad P(Z_t^{(i)} \in \Gamma) = \Psi_t(i, \cdot)U(\Gamma),$$

*where $\Psi_t(i, \cdot)$ is the $i$th row of $\Psi_t$. In (15), $\Psi_t(0, \cdot)$ consists of $\psi_0$ appended with zeros for the continuation states, and the entries for the other rows of $\Psi_0$ are all zero. Also, $\Psi_t$ is calculated from the following iterative scheme, for $t = 1, \ldots, T^*$:*

$$(16) \qquad \Psi_t = \Psi_{t-1}M_t$$

$$\Psi_t(i, (s, \rightarrow)) \leftarrow \Psi_t(i, (s, \rightarrow)) + \Psi_t(i-1, \Gamma) - \Psi_{t-1}(i-1, \Gamma),$$

$$(17)$$

$$i = 2, \ldots, r,$$

*where $\leftarrow$ denotes that the quantity on the right replaces that on the left.*

The theorem is similar to (13), but is modified slightly. The justification for the algorithm (16)–(17) is as follows. Equation (16) is the standard matrix multiplication for each time step as in (13). The extra operation (17) is inserted to increment the row probabilities, as mentioned above, as the elements of row $i$ of $\Psi_t$ give the joint probability of there having been $(i-1)$ occurrences with the chain lying in the various states. With this setup, the probability of being in the absorbing state of $Z_t^{(i)}$ equals the probability of at least $i$ occurrences having taken place. Once the system is evaluated for all $t$ up to $T^*$, the values in the column of $\Psi_{T^*}$ corresponding to the absorbing state $\Gamma$ give the probabilities of the pattern having occurred at least $1, \ldots, r$ times by time $T^*$. More details on the derivation of the algorithm are given in the Appendix.

Theorem 3.3 may be modified to allow for general numbers of occurrences of runs to be counted when using overlapping or nonoverlapping counting (for length exactly $k$ rather than length at least $k$) in a similar manner. The continuation states would not be needed in these cases, as the initial states for the chains $\{Z_t^{(i)}\}$, $i = 2, \ldots, r$ will already be in $S_Z$. As can be seen, when $r = 1$, the iterative scheme of Theorem 3.3 is not needed.

**4. Examples.** In this section two applications of the methodology outlined in the last section are given. The first, to eruptions of the old faithful geyser, is a simple example given mainly to illustrate various aspects of the method, such as the choice of the auxiliary Markov chain, the determination of its state space and the formation of its transition probability matrix. The second example is more substantive, dealing with the $CpG$ island problem that was discussed briefly in the Introduction.

4.1. *Old Faithful geyser.* The durations of successive eruptions from August 1 to August 15, 1985 of the Old Faithful geyser in Yellowstone National Park were presented by Azzalini and Bowman (1990) and analyzed using a second-order Markov chain. The durations were classified as either being short (0) or long (1) to give a binary time series of length $T = 299$. However, what is observed is not necessarily what is happening with the underlying system. Specifically, there could be a long eruption even though the state of the geyser is more in line with a short eruption. MacDonald and Zucchini (1997) extended the model to a second-order HMM to mitigate the noise in the system. Their model is used here to find distributions of patterns in the underlying system.

The state spaces are $S_X = S_Y = \{0, 1\}$. The transition probabilities for the hidden states $\{X_t\}$ are $p(0|0,0) = p(0|1,0) = 0$, $p(0|0,1) = 0.717$, $p(0|1,1) = 0.414$ and $p(1|x_{t-1}, x_t) = 1 - p(0|x_{t-1}, x_t)$. The initial distribution is $\pi(0,0) = 0$, $\pi(1,0) = \pi(0,1) = \pi(1,1) = \frac{1}{3}$. The output probabilities conditioned on states are $\gamma(0|0) = 0.928$, $\gamma(0|1) = 0$, with $\gamma(1|x_t) = 1 - \gamma(0|x_t)$.

This is actually an example of a reduced second-order Markovian state sequence, as $X_{t+1} = 1$ whenever $X_t = 0$, regardless of any values of previous states. This indicates that a short eruption state is always followed by a long one. It is only when the current state is 1 that $X_{t-1}$ is needed to determine probabilities for transitions to $X_{t+1}$. The initial distribution accounts for the fact that the state where $X_{-1} = X_0 = 0$ is not possible under the model.

Two types of patterns in the eruption process are typical; either there are runs in the underlying state process of long eruptions, or the system alternates between the long and short eruption states. Little is known about why these patterns occur other than that the short eruption state is always followed by the long one [Perkins (1997)]. It is of interest to see how these pattern distributions modeled through HMMs correspond to pattern distributions unconditional on data. Here, the alternating pattern is analyzed. The corresponding compound pattern is

$$\Lambda(k) = \overbrace{0101..01}^{k} \cup \overbrace{1010...10}^{k},$$

where $\Lambda(k)$ denotes an alternating run of length exactly $k$, and $\overbrace{\therefore}^{k}$ indicates there are $k$ symbols under the bracket. The above formula represents the case where $k$ is even. If $k$ were odd, the next alternating symbol would be added to each of the two simple patterns of $\Lambda(k)$, that is, the last symbol would be the same as the first. Progress into the simple patterns can be any of the following (assuming that $k$ is even):

$$\{0, 1, 01, 10, 010, 101, \ldots \overbrace{0101..0}^{k-1}, \overbrace{1010...1}^{k-1}\}.$$

The states of $S_Z$ are defined as ordered vector pairs, the first element being the 2-tuple that gives the last two values of the $\{X_t\}$ sequence, and the second element being the progress into a pattern:

$$(00,0),(11,1),(01,01),(01,101),(01,0101),\ldots,(01,\overbrace{1010..1}^{k-1}),$$

$$(10,10),(10,010),(10,1010),\ldots,(10,\overbrace{0101..0}^{k-1}),\Gamma.$$

The states

$$(00,\varnothing),(01,\varnothing),(10,\varnothing),(11,\varnothing),(01,1),(10,0)$$

are also needed; the first four as possible states for $Z_0$, and the latter two as possible values of $Z_1$. Here $\varnothing$ denotes that there is no progress into a simple pattern of the system. These initialization states are needed for times $t < 2$ since, for these values of $t$, some or all of the 2-tuple does not count, rendering pattern progress different than for times $t \geq 2$. After time $t = 1$, the Markov chain can never return to these states, and thus, they are deleted from the state space after that time.

The transition probability matrix for $Z_t$ is constructed using transition probabilities determined through Lemma 3.1. The transition probabilities required are based on the transition probabilities determined for the $\{X_t\}$ sequence through the 2-tuple, while the entire vector state from $S_Z$ determines the possible destination of the state. A few example transition probabilities are listed below, with the others determined in a similar manner:

$$P(Z_t = (10,\overbrace{01\ldots010}^{i+1})|Z_{t-1} = (01,\overbrace{01\ldots01}^{i}))$$
$$= P(X_t = 0|X_{t-2} = 0, X_{t-1} = 1, Y^{(T)}), \qquad i = 1,\ldots,k-2,$$

$$P(Z_t = (11,1)|Z_{t-1} = (01,\overbrace{10\ldots101}^{i}))$$
$$(18) \qquad = P(X_t = 1|X_{t-2} = 0, X_{t-1} = 1, Y^{(T)}), \qquad i = 1,\ldots,k-1,$$

$$P(Z_t = (10,10)|Z_{t-1} = (11,1))$$
$$= P(X_t = 0|X_{t-2} = 1, X_{t-1} = 1, Y^{(T)}),$$

$$P(Z_t = \Gamma|Z_{t-1} = (10,\overbrace{0101\ldots0}^{k-1}))$$
$$= P(X_t = 1|X_{t-2} = 1, X_{t-1} = 0, Y^{(T)}),$$

$$P(Z_t = \Gamma|Z_{t-1} = \Gamma) = 1.$$

Using $\{Z_t\}$, the waiting time probability for an alternating sequence of length $k$ can be found through Theorem 3.2 and then using (14). The probability mass function $P(R(T) = k)$ for alternating runs of lengths $k = 1, \ldots, 70$ given all the data ($T = T^* = 299$) is shown in Figure 1. This plot indicates that, conditional on the observations, only very specific alternating run lengths are likely. Indeed, as the probability of two short eruption states following one another is zero, the pattern is highly likely to start in state 1 and end in state 1, hence, the specific lengths. However, the conditional distribution given the observations is very different from the unconditional distribution, showing that information from the observation sequence is very useful for determining probabilities for patterns in the underlying states.

4.2. *DNA sequence analysis—CpG islands.* The use of HMMs to model DNA sequences with heterogenous segments was pioneered by Churchill (1989), and since that time their use for that cause has increased. As mentioned in the introduction, HMMs have been shown to be especially suitable for the analysis of *CpG* islands.

Define $S_X = \{A^+, C^+, G^+, T^+, A^-, C^-, G^-, T^-\}$, where a superscript "$+$" indicates that the state is within a *CpG* island and "$-$" that it is not, and let $S_Y = \{A, C, G, T\}$. The transition probability matrix associated with the state sequence $\{X_t\}$ is taken to be

$$
(19) \quad
\{p(x_{i+1}|x_i)\} =
\begin{array}{c}
A^+ \\ C^+ \\ G^+ \\ T^+ \\ A^- \\ C^- \\ G^- \\ T^-
\end{array}
\begin{pmatrix}
0.17635 & 0.26838 & 0.41719 & 0.11761 & 0.0037214 & 0.0055995 & 0.0086354 & 0.0025226 \\
0.16737 & 0.36005 & 0.26811 & 0.18400 & 0.0035381 & 0.0074703 & 0.0055940 & 0.0038774 \\
0.15775 & 0.33201 & 0.36726 & 0.12250 & 0.0033417 & 0.0068982 & 0.0076165 & 0.0026225 \\
0.077468 & 0.34768 & 0.37607 & 0.17831 & 0.0017034 & 0.0072179 & 0.0077973 & 0.0037613 \\
0.0004247 & 0.0003297 & 0.0004087 & 0.0003347 & 0.29953 & 0.20472 & 0.28456 & 0.20971 \\
0.0004466 & 0.0004227 & 0.0002019 & 0.0004266 & 0.32148 & 0.29753 & 0.077969 & 0.30152 \\
0.0003018 & 0.0003637 & 0.0004167 & 0.0004167 & 0.17677 & 0.23865 & 0.29154 & 0.29154 \\
0.0003727 & 0.0003707 & 0.0004227 & 0.0003327 & 0.24763 & 0.24563 & 0.29753 & 0.20771
\end{pmatrix}.
$$

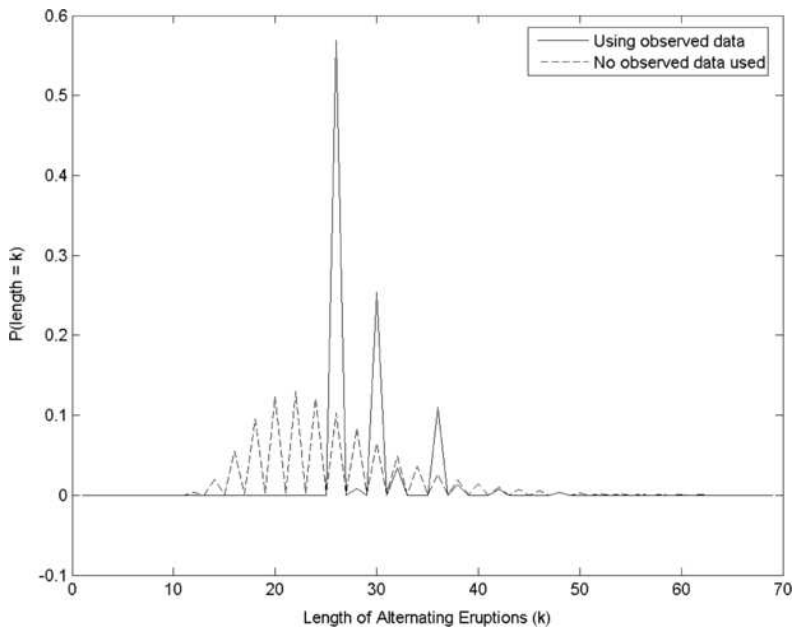

FIG. 1. *Graph of the probability mass function of the length of the longest alternating underlying short/long eruption states, both conditional and unconditional on the observation sequence, for the Old Faithful geyser, Yellowstone National Park, from August 1 to August 15, 1985.*

This matrix is based on the transition probability matrices given in Durbin, Eddy, Krogh and Mitchison (1998), which were calculated using maximum likelihood methods from human DNA with 48 putative $CpG$ islands present (which were predetermined using other methods).

If $Q$ represents a generic nucleotide, that is, $Q \in \{A, C, G, T\}$, then $\gamma(Q|Q^-) = \gamma(Q|Q^+) = 1$. Even though the observed nucleotide is totally determined if the underlying state is known, it is not possible to know whether an observation came from a $CpG$ island or not. Finally, define $\pi(A^-) = \pi(C^-) = \pi(G^-) = \pi(T^-) = \frac{1}{4}$, that is, the underlying state sequence is equally likely to start in any of the non-$CpG$ island states.

The patterns of interest are sequences of $+$'s, which indicate that the system is currently in a $CpG$ island (since $+$ denotes $A^+ \cup C^+ \cup G^+ \cup T^+$). This is slightly different from the traditional definition of runs (consecutive occurrences of one symbol). Of interest will be the distribution of the length of $CpG$ islands in a given sequence of nucleotides, and the distribution of the number of $CpG$ islands of length at least $k$, for specified values of $k$.

The state space for the auxiliary $\{Z_t\}$ chain is given by

$$
(20) \quad \bigcup_{Q \in \{A,C,G,T\}} \{(Q^+, +), (Q^+, ++), \ldots, (Q^+, \overbrace{+\cdots+}^{k-1}), Q^-\} \cup \Gamma,
$$

where $k$ is a specified run length, $(Q^+, \overbrace{+\cdots+}^{i})$ gives the value of $X_t$ and the current run of $+$'s, and $\Gamma$ is an absorbing state to indicate that $k$ consecutive $+$'s have occurred. The desired run occurs by time $t$ if and only if $Z_t = \Gamma$, and thus theorems from Section 3 may be used to compute probabilities. As it is assumed that the state sequence does not begin in a $CpG$ island state, no additional states are needed to initialize the chain. [To relax this assumption, the states $\bigcup_{Q \in \{A,C,G,T\}} \{(Q^+, \varnothing)\}$ would need to be added as initialization states for $Z_0$.]

The HMM described in this subsection was used to analyze several human genomic sequences. First, chromosome 20, locus AL133339 [Barlow (2005)] with 18,058 base pairs was investigated (thus, $T = T^* = 18{,}058$), that is, AL133339 is the observed sequence $\{Y_t\}$. This sequence is known to contain at least one example of a putative $CpG$ island. Given that the definition of a $CpG$ island is slightly open to interpretation, the distribution of the $CpG$ island lengths in the sequence is of interest. This is calculated through (14), if $\Lambda(k)$ is taken to mean a run of $+$'s of length $k$. Waiting time probabilities are computed through Theorem 3.2.

Approximate locations of $CpG$ islands are indicated by the jumps in the probability plot of Figure 2(a). In that plot, when $k = 100$, there is a jump around sequence position 4,500, and then another jump approximately at position 14,000. However, when the length $k$ is increased, the jump around sequence position 4,500 disappears, indicating that the first $CpG$ island is smaller than the second one. These locations and size differences were verified by using the software $CpG$ islands searcher [Takai and Jones (2003)], which is based on the (non-HMM) algorithm of Takai and Jones (2002) and requires the use of predetermined thresholds. However, though the analysis using the latter software concurred with the relative lengths found in this work, the actual lengths of the $CpG$ islands given by the software (which gives a fixed length and not a distribution) were longer than those expected using the full distribution. Note that both the $CpG$ island lengths determined by the software, and the length distribution computed here, depend on initial settings, the initial parameters for the software or the transition probabilities for the $\{X_t\}$ sequence when using HMMs.

The standard method of detecting $CpG$ islands using HMMs is to use the Viterbi algorithm to find the most probable state sequence and then search for $CpG$ islands as though the derived state sequence gives the true underlying states. However, as was mentioned in the introduction, although the Viterbi algorithm gives the most probable state sequence, using it does not necessarily lead to the most probable $CpG$ island length. Using the Viterbi algorithm, a $CpG$ island of length 362 is found in the data. This corresponds to the small jump in the distribution at the corresponding location [toward the right tail of the probability distribution plotted in Figure 2(b)]. Thus, while 362 is the $CpG$ island length from the most probable state sequence, it is not the most probable $CpG$ island length for this data set.

By definition, $CpG$ islands are often required to be at least 100 base pairs long, and thus, the number of islands of at least size 100 is of significant interest. Defining continuation states $\bigcup_{Q \in A,C,G,T} \{(Q^+, \rightarrow)\}$, and using Theorem 3.3, the distribution of the number of $CpG$ islands of length at least $k$ is computed (Figure 3). This distribution yields an estimated mean for the number of islands of length at least 100 in the sequence of 3.26, and a mode of 3 islands. Using the method based on the Viterbi algorithm [Durbin, Eddy, Krogh and Mitchison (1998)], four islands were identified, however, only two of them are at least 100 base pairs in length, showing

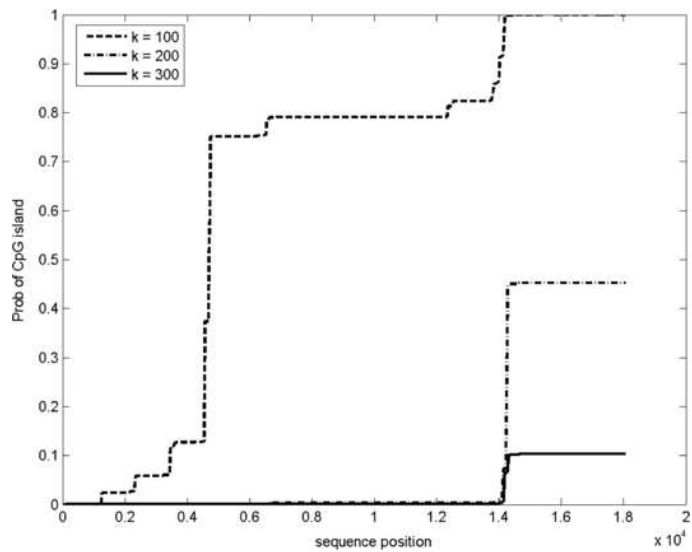

(a) Probability distribution of a *CpG* island of length at least *k* having occurred at or before different sequence positions.

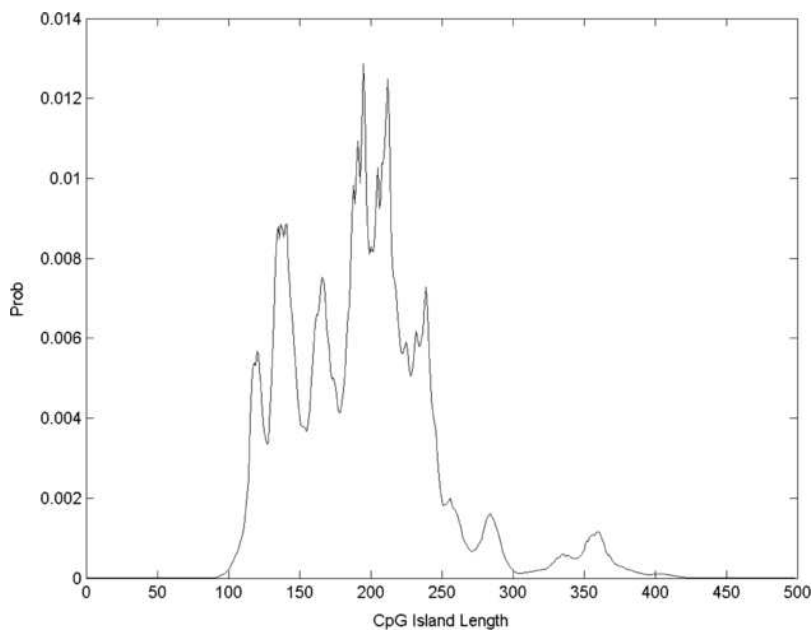

(b) Probability mass function of maximal *CpG* island length.

FIG. 2. *CpG island distributions for the 18,058 nucleotide gene sequence locus AL133339 from Human Chromosome 20. The first plot shows the probability of islands of lengths k = 100, k = 200, and k = 300 occurring at or before the various locations of the sequence, while the second gives the probability mass function of the maximal CpG island length using the entire data sequence.*

that the most probable state sequence does not necessarily give the most probable number of islands when important biological constraints are applied.

It would be very difficult to find a way to constrain the Viterbi algorithm to find the most probable state sequence subject to biological constraints. However, if, once the most likely state sequence is found using the Viterbi algorithm, biological constraints are imposed, the state sequence is modified, changing some island states into non-island states. The state sequence is then not guaranteed to be the most probable one, nor even the most probable state sequence among those which conform to the desired constraints.

For the purposes of counting, it is often desirable to have gaps of at least a certain length, before allowing a new island to start. This is biologically realistic as islands are presumed not only to be longer than a certain length but also to be infrequent and not particularly close to one another [Takai and Jones (2002)]. In this case, a problem again arises when using the Viterbi algorithm for $CpG$ island detection. If two islands are close together, they are deemed one island and again the state sequence found by the Viterbi algorithm is modified (the intermediate non-island states in the sequence between the two islands are changed to be island states). This also cannot then be guaranteed to be the most probable state sequence among all sequences that conform to the biological constraints. On the other hand, gaps are very easily handled in the computation of pattern distributions presented here.

The distribution of the number of $CpG$ islands of at least a certain length, and having gaps separating them, may be calculated by extending the ideas of ordered patterns [Fu and Chang (2003)]. However, the state space needed would become very large. It is much more efficient to calculate the distribution using an argument very similar to that used to compute the distribution of the total number of islands of length at least 100 (Theorem 3.3). In this case, two Markovian systems are used, one to find $CpG$ islands of at least a certain length, and another to find non-$CpG$ island intervals after the $CpG$ island has finished. This essentially requires finding patterns of +'s in one conditional Markov chain and then patterns of −'s in another conditional chain. The probability of entering the absorbing state of a "+" pattern is then used as the probability for an initial state of the conditional chain for the "−" pattern and visa versa, in a similar manner as the absorption probability from the $(i-1)$st chain is used as the initial probability in (23) of the Appendix. For example, the second "+" chain is now conditioned on both a "+" pattern having occurred previously and also a separation of "−" of a certain length having occurred after that.

A separation of at least 100 base pairs is often chosen to be the minimum that is required for two $CpG$ islands to be distinct [Takai and Jones (2003)]. The $CpG$ island searcher software requires a minimum of 200 base pairs to

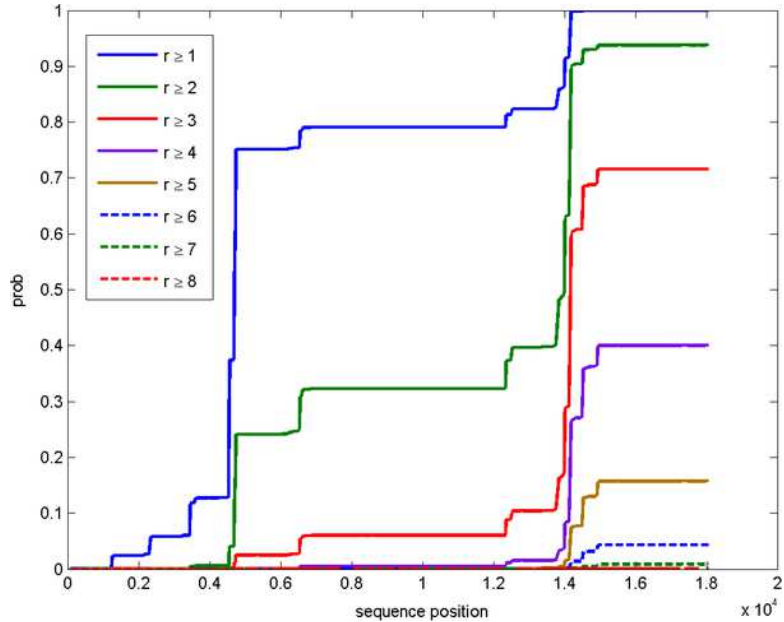

(a) Probability distribution of number $r$ of $CpG$ islands of length at least 100 having occurred at or before different sequence positions.

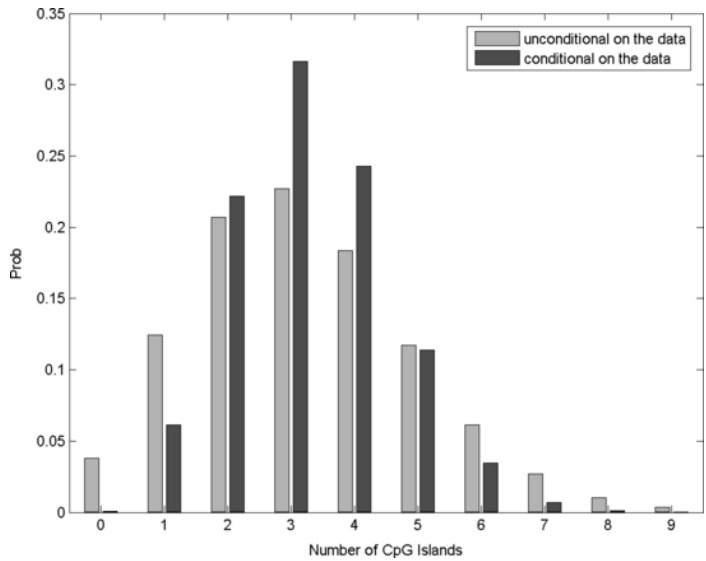

(b) Probability mass function of number of $CpG$ islands of length at least 100.

FIG. 3. *CpG island distributions for the 18,058 nucleotide gene sequence locus AL133339 from Human Chromosome 20. The first plot shows the probability of having r islands of lengths at least 100 occur at or before the various locations of the sequence. The second plot depicts the probability mass function of the number of CpG islands of length at least 100 using the entire observed data sequence, as well as unconditional on the data.*

be present for a segment to be counted as a *CpG* island, as well as a minimum separation of at least 100 base pairs between islands, and these lengths will be used here to provide a comparison. However, the analysis could easily be modified to deal with other biological constraints, if necessary. The *CpG* islands present in the DNA sequence AL133339 are relatively short (as evidenced by the length distribution), which means that there is less chance of the patterns being split into two islands with both being counted as being above a certain length. However, the DNA sequence AL117335 [Smith (2007)] of length 44,527 base pairs, also from chromosome 20, contains an example of a long *CpG* island. If this island were split without a minimum separation distance between the pieces, a higher number of islands could be predicted by the model. Indeed, when using the Viterbi algorithm, it is not unusual to see islands with small breaks between them in the most probable state sequence (without any constraints placed on it). These would then be modified and counted as being from one island. Thus, the condition of a minimum separation between islands is a useful biological constraint that should be imposed on the analysis.

As expected, adding the biological restriction of a minimum gap length of 100 base pairs reduces the likely number of islands counted in the data [see Figure 4]. The corresponding probabilities of occurrences at or before certain sequence locations under this restriction can also be found and are given in Figure 4(a). Using the Viterbi algorithm, seven islands in total are found in AL117335. However, only two of those have more than 200 base pairs and a separation of more than 100 base pairs. There are two additional islands which are close to being counted (one island which, although longer than 200 base pairs, is only 95 base pairs from another, and another one that is only 193 base pairs long). By using the *CpG* island searcher software with the same restrictions, three islands were found in the sequence. As can be seen in Figure 4(b), it is likely that there are two or three islands in the data, probably arising close to locations 2,000, 18,000 and 26,000 [see Figure 4(a)]. While the first island was not confirmed by the *CpG* island searcher software, the locations of the latter two islands were. Thus, by using a combination of the distribution estimates and the probabilistic plots, an additional possible location for a *CpG* island has been identified, which, although not satisfying the predetermined threshold requirements used in the software, is consistent with the HMM that is being used in the present analysis to identify *CpG* islands.

One could compute probabilities for the length of combined islands by allowing the desired pattern to be a *scan*, thus extending work for scans in Markovian sequences [see, e.g., Naus (1982), Koutras and Alexandrou (1995)] to the HMM setting. However, careful definitions of what constitutes a scan would need to be made so that the output would constitute a biologically plausible island.

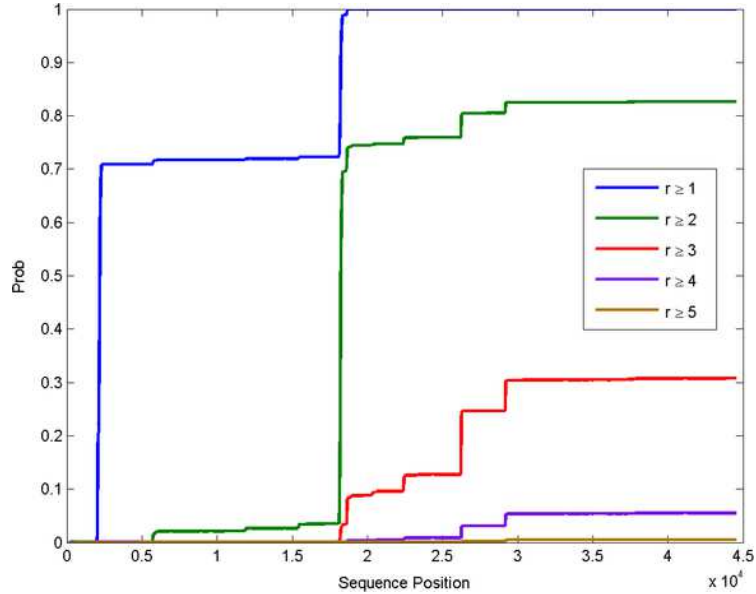

(a) Probability distribution of number $r$ of *CpG* islands of length at least 200 with a separation gap of at least 100 base pairs having occurred at or before different sequence positions.

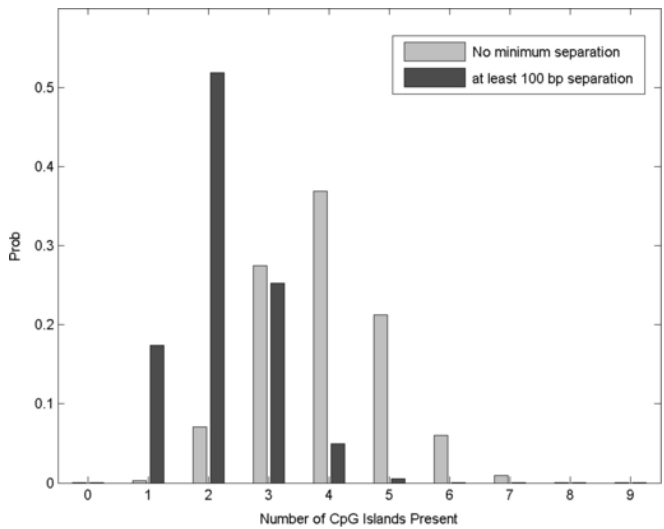

(b) Probability mass function of number of *CpG* islands of length at least 200, with and without a separation gap of 100 base pairs required.

FIG. 4. *CpG island distributions for the 44,527 nucleotide gene sequence locus AL117335 from Human Chromosome 20. The first plot shows the probability of the number of islands $r$ of length at least 200 occurring at or before the various locations of the sequence, while the second gives the probability mass function of the number of CpG islands of length at least 200 using the entire data sequence. The requirement that there should be a gap between CpG islands of at least 100 base pairs is accounted for in the distributions.*

Computing these distributions can be computationally expensive in terms of CPU time. The programs were written in MATLAB and run on a standard desktop PC. To run the example for AL133339 with over 18,000 base pairs to find islands of length at least 100 takes approximately 8 minutes on a single PC. However, the actual time used in matrix multiplication is only 5 seconds, with the need to constantly reassign the inhomogeneous transition probabilities to locations in the transition probability matrix requiring the rest of the time. However, this means that considerable savings can be made when determining the length distribution by calculating it at once rather than sequentially, as the submatrices of a large transition matrix can be used to calculate distributions at each time point without needing to reassign probabilities.

The calculation for AL117335 with over 44,000 base pairs, islands of at least length 200 and separation gaps of least 100 base pairs takes just over 40 minutes on a single PC. The use of a PC cluster could certainly facilitate much quicker computation of these distributions.

**5. Discussion.** Distributions associated with runs and patterns in random sequences are usually computed based on some model, but with no additional information. Here, a method is introduced to compute such probabilities in situations where $T$ observations of an observed sequence have been realized, but where the patterns of interest occur in an underlying state sequence. The observation sequence gives incomplete or noisy information concerning the values of the unobserved sequence. A hidden Markov model of an arbitrary order $m$ is used to model this situation. Using the methodology of this paper, distributions associated with patterns in an underlying state sequence may be computed either up to index $T$, or to an index $T^* > T$.

Probabilities are computed by associating with the underlying state sequence a Markov chain that lies in its absorbing state if and only if the pattern of interest has occurred. This allows the computation of probabilities through basic properties of Markov chains. The types of patterns that may be investigated are of a general nature and the distributions can be found for a variety of features, including, but not limited to, the waiting time for a pattern to occur, the maximal length of a run or repeated pattern and the number of pattern occurrences, all conditional on the observed data and possibly subject to additional constraints. As has been seen, these features have direct relevance to many applications.

The computations themselves may be quite intensive, but are made simpler by the $m$th order forward-backward algorithm, which provides a quick method to determine the transition probabilities for the underlying state sequence, and thus also for the auxiliary Markov chain. The number of additional states needed in $S_Z$ over those required for $S_X$ depends on several factors. While the worst case scenario is that the number of additional states increases exponentially with the order of Markovian dependence, patterns are often associated with only a small number of states of $S_X$, and the number of additional states required then increases exponentially in only these states. This is especially true in cases such as simple patterns or small compound patterns, as in the examples. In simple patterns, for example, runs of one symbol, only one state is associated with the pattern, and thus, the number of additional states does not increase at all. The complexity also increases with the length of pattern, the number of patterns to be found and the length of the data, but the increases are only of linear order. It should also be noted that the structure of these transition matrices can become incredibly sparse, which aids computation considerably. As an example, if the order of the underlying Markovian sequence was changed to second order in the $CpG$ islands example, approximately four times as many states would be needed for each computation, although most of these will have transition probabilities of zero between them.

In practice, the slowest part of computation often arises from the need to continually reassign the transition probabilities to appropriate locations in the transition probability matrix, due to the inhomogeneous nature of the auxiliary Markov chain. The transition probabilities can all be calculated quickly outside the algorithm, but assigning them to the correct positions in the $M_t$ matrix can be time consuming. This might be overcome, however, by using computer languages that specialize in this sort of operation. However, in order to circumvent this problem when calculating distributions of run lengths, the distributions can be calculated at each time point using submatrices of a large transition matrix, rather than doing it sequentially.

An alternative method for calculating the distributions is to sample from the conditional distribution of the states through the conditional transition probabilities given by Lemma 3.1 and then empirically calculate the distributions of pattern lengths and number of patterns from these samples. For especially complex patterns, this could well lead to improvements in the computation time, although the computed distribution would then be approximate. However, given the inhomogeneous nature of the transition probabilities and the large number of samples needed to get accurate distributions for long sequences, unless the patterns are quite complicated, this method is likely to be less efficient. For example, to estimate distributions of $CpG$ islands of length at least 100, the computational cost of generating one sample from the distribution (requiring 64 transition probabilities at each point) is 1/50 of the cost of finding the exact distribution (requiring 3217 transition probabilities at each time point). Thus, only 50 samples could be drawn before the sampling method is more computationally expensive. This ratio obviously decreases as the length of the pattern increases, but only linearly (e.g., to find the exact distribution of islands of length at least 500 requires approximately 16,000 transition probabilities at each point). While the overhead of assigning the probabilities to the transition matrices have not been included in this calculation, the storage requirements for generating large numbers of samples and then searching these samples have also not been included. These things being considered, as the patterns get more complicated in terms of number or length, computation through sampling will become relatively more efficient, and may well be appropriate in some applications, especially for finding the distribution of the longest run. In the examples here, using the sampling methods would not be computationally advantageous, and, as they also do not lead to exact distributions, they were not used.

While the results of this paper facilitate the computation of general waiting time probabilities, other possible applications of the methodology exist, an example being change point problems. Indeed, the $CpG$ island example above could be considered as a problem of detecting change points between $CpG$ islands and non-$CpG$ islands with Figures 2(a)–4(a) giving information about the locations of those change points.

Two issues have not been addressed in this paper. First, in many instances, the transition probabilities of the underlying state sequence need to be estimated, not an easy task since the sequence is hidden. There has been considerable work on this problem [see Cappé, Moulines and Rydén (2005) for the latest in techniques in this area]. In this work, it is assumed that appropriate values for the transition probabilities have been estimated, with the estimated parameters being used as input to compute the desired waiting time distributions. There could, however, be an appreciable effect of the estimation process on the probabilities that are eventually determined. An area of future work is to determine the effect of small errors in parameter estimation on the output waiting time distributions for patterns.

Second, in this paper the assumption is made that the length $T$ of the observation sequence is fixed throughout the experiment. In many applications, such as for DNA sequences, this assumption is appropriate, but in others, observations may come in a continual stream, yielding a sequential analysis problem. While, in principle, there is no restriction on using the methods of this paper to update distributions when a new observation arrives, computationally, this task may be very burdensome. New transition matrices for the underlying state sequence would be needed after every new observation, not only for future points but also for those in the past as well, as the backward probabilities would change for all $t$. Whether approximations can be found to ease this burden will also be a topic for future consideration.

It is well known that one of the main problems associated with higher-order HMMs, and higher-order Markov chains in general, is the growing number of parameters as the order of dependence increases. This is one of the reasons why higher-order HMMs are not used more frequently. However, the methods presented here may be applied to various forms of higher-order HMMs, including models with reduced parameter sets, such as that proposed by Raftery (1985). Indeed, the Old Faithful geyser example demonstrates a second-order HMM with a reduced parameter set.

In conclusion, this paper has suggested a general approach to the computation of waiting time distributions for patterns in states of HMMs that is easy to implement, and yet also flexible enough to be applicable to a wide variety of applications in science and engineering.

## APPENDIX A: DERIVATION OF ALGORITHM

Let $\phi_{0,t}^{(i)}$ be the initial distribution for the Markov chain associated with the $i$th occurrence, $i = 2, 3, \ldots, r$, when the chain starts at time $t$. Since $i > 1$, the chain will start in one of the continuation states. Define $W\{\Lambda^i(k)\}$ to be the waiting time for the $i$th occurrence of the pattern. Then

$$
\begin{aligned}
(21) \quad & P(W\{\Lambda^i(k)\} \leq t) \\
& = \sum_{0 \leq t_1 \leq t} P(W\{\Lambda^{i-1}(k)\} = t_1) \\
& \quad \times P(W\{\Lambda^i(k)\} \leq t | W\{\Lambda^{i-1}(k)\} = t_1).
\end{aligned}
$$

This can be simplified using a renewal argument. The second term in the summation in (21) is the same as $P(W\{\Lambda(k)\} \leq (t - t_1))$, assuming that the chain starts at time $t_1$ (the chain is inhomogeneous, and thus, the starting time is needed), as the probability of the $i$th occurrence conditioned on the $(i-1)$st can be calculated using the same probability structures as those of the first occurrence (but starting in one of the continuation states). Hence, (21) equals

$$
\begin{aligned}
(22) \quad & \sum_{0 \leq t_1 \leq t} P(W\{\Lambda^{i-1}(k)\} = t_1) P(W\{\Lambda(k)\} \leq (t - t_1)) \\
& = \sum_{0 \leq t_1 \leq t} P(W\{\Lambda^{i-1}(k)\} = t_1) \phi_{0,t_1}^{(i)} \left( \prod_{j=t_1+1}^{t} M_j \right) U(\Gamma)
\end{aligned}
$$

by Theorem 3.2, where the product is taken to be an identity matrix if $t_1 = t$.

Define

$$
\begin{aligned}
(23) \quad \psi_{0,t_1}^{(i)} & = P(W\{\Lambda^{i-1}(k)\} = t_1) \phi_{0,t_1}^{(i)} \\
& = [P(W\{\Lambda^{i-1}(k)\} \leq t_1) \\
& \quad - P(W\{\Lambda^{i-1}(k)\} \leq t_1 - 1)] \phi_{0,t_1}^{(i)}.
\end{aligned}
$$

Then (23) equals

$$(24) \qquad \sum_{0 \leq t_1 \leq t} \psi_{0,t_1}^{(i)} \left( \prod_{j=t_1+1}^{t} M_j \right) U(\Gamma),$$

which when expanded out equals

$$(25) \qquad \overbrace{(\ldots(\psi_{0,0}^{(i)} M_1 + \psi_{0,1}^{(i)}) M_2 + \cdots + \psi_{0,t-2}^{(i)}) M_{t-1} + \psi_{0,t-1}^{(i)}) M_t + \psi_{0,t}^{(i)}}^{t} U(\Gamma),$$

where the probability of seeing the $(i-1)$st occurrence at time 0 is defined to be 0 and hence, $\psi_{0,0}^{(i)} = 0$. If $\psi_{0,0}^{(1)} \equiv \psi_0$ and $\psi_{0,t}^{(1)} \equiv 0, t > 0$, then the algorithm given in (16)–(17) iteratively calculates the set of sums and products of each bracket given in (25) for all $i = 1, \ldots, r$ occurrences simultaneously in $i$ before the multiplication with $U(\Gamma)$.

In the above it has been implicitly assumed that only one state leads to absorption, as in the $CpG$ island example (the deterministic output $Y_t$ guarantees this), and as such $\phi_{0,t_1}^{(i)}$ is a unit vector with a 1 in the initial state. However, the algorithm can be easily modified in the case where there are multiple states leading to absorption. This modification is needed, for example, in the case when calculating $CpG$ island distributions unconditional on the data, where four states can lead to absorption.

**Acknowledgments.** The authors would like to express their sincere thanks to Professor Cheng-Der Fuh for his careful reading of an earlier version of the manuscript and his very helpful suggestions. The authors would also like to express their gratitude to the editor, the associate editor and the referee, whose suggestions helped greatly to improve the paper.

INSTITUTE OF STATISTICAL SCIENCE
ACADEMIA SINICA
128 ACADEMIA ROAD, SEC 2
TAIPEI 11529
TAIWAN
ROC
E-MAIL: jaston@stat.sinica.edu.tw

DEPARTMENT OF STATISTICS
NORTH CAROLINA STATE UNIVERSITY
CAMPUS BOX 8203
2501 FOUNDERS DRIVE
RALEIGH, NORTH CAROLINA 27695-8203
USA
E-MAIL: martin@stat.ncsu.edu